# Prediction of anomalous LA-TA splitting in electrides

Leilei Zhang,[1] Hua Y. Geng,[1,*] and Q. Wu[1]

[1]*National Key Laboratory of Shock Wave and Detonation Physics, Institute of Fluid Physics, CAEP, P.O. Box 919-102, Mianyang, Sichuan 621900, P.R. China*

**Abstract**: Electrides are an emerging class of materials with excess electrons localized in interstices and acting as anionic interstitial quasi-atoms (ISQs). The spatial ion-electron separation means that electrides can be treated physically as ionic crystals, and this unusual behavior leads to extraordinary physical and chemical phenomena. Here, a completely different effect in electrides is predicted. By recognizing the long-range Coulomb interactions between matrix atoms and ISQs that are unique in electrides, a nonanalytic correction to the forces exerted on matrix atoms is proposed. This correction gives rise to an LA-TA splitting in the acoustic branch of lattice phonons near the zone center, similar to the well-known LO-TO splitting in the phonon spectra of ionic compounds. The factors that govern this splitting are investigated, with isotropic fcc-Li and anisotropic hP4-Na as the typical examples. It is found that not all electrides can induce a detectable splitting, and criteria are given for this type of splitting. The present prediction unveils the rich phenomena in electrides and could lead to unprecedented applications.

**I. Introduction**

The phonon dispersion relation $\omega(\mathbf{q})$ of a crystalline lattice reflects the energy variation in quantum vibration of a solid. This quantity encodes all information about the ionic contributions to the thermodynamics properties and dynamic process at low temperature, which is fundamental for both equilibrium and non-equilibrium statistics of a solid. Normally, there are 3 acoustic branches and (3$N$-3) optical branches for a 3-D crystal composed by $N$ ($N\rightarrow\infty$) atoms. Their frequencies are denoted as $\omega_A$ and $\omega_O$, respectively. For non-polar crystals, the long wavelength longitudinal optical (LO) and transverse optical (TO) mode could be degenerate nears the zone center (ZC). However, in ionic crystals, the relative vibrations of positively and negatively charged ions induce macroscopic electric field (E-field), which in turn affects the motion of ions [1,2]. This long-range Coulomb interaction (LRCI) lifts the aforementioned degenerate, and gives rise to the LO-TO splitting near the Brillouin ZC ($\omega_{LO}\neq\omega_{TO}$, $q\approx 0$), which is very important for the accurate description of dielectric properties and phonon transportation in a polar system, and have been widely explored [3-8].

[*] To whom correspondence should be addressed. E-mail: s102genghy@caep.cn

Electrides [9-20] are an emerging class of materials in which highly-localized excess electrons locate at lattice interstitial sites and behave as anionic interstitial quasi-atoms (ISQ) [11,21-24]. Electrides have promising physical and chemical properties due to the unique electronic structures that originated from the localized ISQs. For example, it open the energy gap and lead to counterintuitive metal-nonmetal transition in Li [25,26] and Na [27], as well as the accompanied complex structural phase transition in dense Li [26]. This extraordinary behavior of the localized electrons not only modifies the electronic and crystalline structure greatly, but also changes the dielectric and optical properties [28], which makes a semi-transparent metal become possible [28].

In physics, electride is analog to ionic compound, and is polar intrinsically. The difference is the negatively charged ISQs take the role of anions in electrides. Therefore, vibrations of positively charged matrix atoms in electride will induce a non-negligible macroscopic E-field, which leads to LRCI and a non-analytical contribution to the dielectric terms near the ZC. In this work, a theoretical model to describe this non-analytical contribution is proposed. The resultant effect in the phonon dispersion is explored, which gives an anomalous splitting in the long wavelength longitudinal acoustic (LA) and transverse acoustic (TA) branches near the ZC. This novel but counterintuitive LA-TA splitting in electrides is a direct counterpart of the well-known LO-TO splitting in classical polar system, and demonstrates the direct physical consequence of the strong long-range coupling of localized electrons and crystalline matrix atoms. It also acts as a typical example that reveals the direct participation of electrons in sound propagation in condensed matters.

In section II, the methodology and computational details are presented. The results and analysis are given in section III. The criterion and the thumb rule for this extraordinary LA-TA splitting to occur are summarized in section IV, together with relevant discussions. Finally, the conclusion and summary of main findings are given in section V.

**II. Methodology and computational details**

**A. Basic theory**

In quantum mechanics for solid, the electronic and nuclear subsystems can be decoupled by using the Born-Oppenheimer approximation, with the nuclear part described by the lattice dynamics. The motion of ions in a crystalline solid shows a periodic pattern, and vibrates around their equilibrium position. The vibrational frequencies $\omega(\mathbf{q})$ are determined to be the square root of eigenvalues of the dynamical matrix $D_{s\alpha,t\beta}(\mathbf{q})$ [29] following the lattice dynamics theory:

$$D_{s\alpha,t\beta}(\mathbf{q}) = \frac{1}{\sqrt{M_s M_t}} \sum_l \Phi_{ls\alpha,0t\beta} \exp[i\mathbf{q}\cdot(\mathbf{R}_0 + \boldsymbol{\tau}_t - \mathbf{R}_l - \boldsymbol{\tau}_s)], \qquad (1)$$

which is the lattice Fourier transformation of the force constants exerting on the ions. Here

$\mathbf{R}_l + \boldsymbol{\tau}_s$ represents the equilibrium position of atom *s* with mass $M_s$ in primitive cell *l*, and the sum runs over the infinite number of primitive cells in the crystal. $\Phi_{ls\alpha,0t\beta}$ is the force-constant matrix, given by the second-order derivatives of potential *U* with respect to the ionic displacement *u*, i.e. $\Phi_{ls\alpha,0t\beta} = \partial^2 U / \partial u_{ls\alpha} \partial u_{0t\beta}$, evaluated with all atoms at their equilibrium positions, and *α* and *β* are Cartesian components.

This standard formulation, however, does not include the non-analytic term derived from long-range interactions, which arises in polar system due to the LRCI[1,2]. A typical example is the ionic crystal, in which the relative displacement of charged cations and anions creates macroscopic E-field, which in turn affects the vibrations of the charged ions. This additional forces acting on ions are non-analytic, and can be expressed as a correction term to the dynamical matrix at small wave-vector **q** near the ZC[29]:

$$D_{s\alpha,t\beta}^{\text{na}}{}^{(\mathbf{q}\to 0)} = \frac{1}{\sqrt{M_s M_t}} \frac{4\pi e^2}{\Omega} \frac{(\mathbf{q}\cdot \mathbf{Z}_s^*)_\alpha (\mathbf{q}\cdot \mathbf{Z}_t^*)_\beta}{\mathbf{q}\cdot \boldsymbol{\varepsilon}^\infty \cdot \mathbf{q}}, \qquad (2)$$

where $\mathbf{Z}_s^*$ is the Born effective charge tensor for atom *s*, $\boldsymbol{\varepsilon}^\infty$ is the high frequency static dielectric tensor, $\Omega$ is the unit cell volume. This non-analytic correction lifts the energy degeneracy and induces the LO-TO splitting around the Γ point in ionic crystal.

Electride is unique in that it is intrinsically polar even in the elemental phases such as that of Li and Na[25-28]. The spatial ion-electron separation and charge transfer in this system leads to dipole moments and the subsequent macroscopic E-field when the matrix atoms move around, very similar to the well-known case of ionic compound. The fact that the highly localized excess electrons in electride form negatively charged ISQs, as well as the observations that ISQs almost remain intact when the matrix atoms vibrate under high-temperature/pressure condations[27,30] and that they could form covalent bonds[31,32], support us to confidently map an electride onto an ionic compound, with the matrix atoms and ISQs correspond to the cations and anions, respectively. Also, this treatment strategy can be regarded as integrating the electronic degree of freedom to the sites and charge of the ISQs, which is a kind of the coarse-graining approach method being widely used in condensed-matter physics. This standard treatment of electride makes it possible to describe the non-analytic contribution of the LRCI in electrides by generalization of Eq.(2) to include ISQ as a new species. The additional components for the non-analytic dynamical matrix thus should be

$$D_{s\alpha,(ISQ)\beta}^{na}{}^{(\mathbf{q}\to 0)} = \frac{1}{\sqrt{M_s M_{ISQ}}} \frac{4\pi e^2}{\Omega} \frac{(\mathbf{q}\cdot\mathbf{Z}_s^*)_\alpha (\mathbf{q}\cdot\mathbf{Z}_{ISQ}^*)_\beta}{\mathbf{q}\cdot\varepsilon^\infty\cdot\mathbf{q}}, \qquad (3)$$

and

$$D_{(ISQ)'\alpha,(ISQ)''\beta}^{na}{}^{(\mathbf{q}\to 0)} = \frac{1}{\sqrt{M_{(ISQ)'} M_{(ISQ)''}}} \frac{4\pi e^2}{\Omega} \frac{(\mathbf{q}\cdot\mathbf{Z}_{(ISQ)'}^*)_\alpha (\mathbf{q}\cdot\mathbf{Z}_{(ISQ)''}^*)_\beta}{\mathbf{q}\cdot\varepsilon^\infty\cdot\mathbf{q}}. \qquad (4)$$

Eq.(1) and Eqs.(2-4) constitute the full dynamical matrix for an electride.

In Eqs.(3) and (4), we introduce the $\mathbf{Z}^*$ for anionic ISQs, which unfortunately cannot be directly evaluated by using current first-principles polarization theory. Nonetheless, they can be estimated with the help of lattice symmetry and the acoustic sum rule of $\sum_k Z_{k,\alpha\beta}^* = 0$. Alternatively, since the Bader charge of the matrix atoms and ISQs can be readily computed using the electron density[33], it is plausible to approximate the $\mathbf{Z}^*$ in Eqs. (2-4) by the Bader charge[34]. As for the relative atomic mass of ISQ ($M_{ISQ}$), we take the approximation of $M_{ISQ}\to\infty$ following the observation that the position of ISQ in Li and Na keeps almost unchanged during the vibrations of matrix atoms. On the other hand, the change of ISQs induced by lattice atoms is similar to the oscillation of plasmon. The motion frequency of plasmon is $\omega = \sqrt{\dfrac{4\pi e^2 n_e}{m^*}}$, where $m^*$ is the effective mass of electrons. For electrides, the valence electrons are highly localized in the interstitial sites, so that its effective mass $m^*$ is very large. As a result, it leads to a relatively small excitation frequency and relatively slow collective motion. Therefore, the vibration of ISQs in electrides can be coupled with matrix atoms. In spite of that, its impact to the phonon dispersions also will be discussed in following sections by using a finite value.

**B. Computational details**

We explored the LRCI effect on phonon dispersions (Eqs. (2-4)) in promising electrides at various pressures, including Li, Na, K, Rb, Cs, Sr, Be, Mg, and Ca. Here, only the results of dense FCC-Li and hP4-Na (see Fig. 1) will be reported. The structural relaxation and force calculation are carried out with density functional theory (DFT)[35,36] as implemented in the Vienna Ab-initio Simulation Package (VASP)[37-39]. The projector-augmented wave (PAW) pseudo-potential[39,40] method and plane-wave basis set are employed. The generalized-gradient approximation (GGA) of Perdew-Burke-Ernzerhof (PBE) is used for the electronic exchange-correlation functional[41]. The $1s^22s^1$ and $2s^22p^63s^1$ electrons are included in the valence space for Li and Na, respectively. The required forces are computed using a 5×5×5 supercell containing 125 atoms with an energy cutoff of 1000 eV in FCC-Li phase and a 4×4×3 supercell containing 192 atoms with an energy cutoff of 1000 eV in hP4-Na phase, respectively. A Γ-centered k-point mesh of 5×5×5 and 3×3×3 are employed to sample the Brillouin zone (BZ). The total energy is ensured to converge to within $1\times10^{-6}$ eV/atom. All investigated structures are fully relaxed at the given pressure until the Hellmann-Feynman forces acting on all atoms less than 0.001 eV/Å, and the total stress tensor is converged to the hydrostatic state within 0.01 GPa. The $\mathbf{Z}^*$ and $\varepsilon^\infty$ are evaluated with density functional perturbation theory method, as proposed by Baroni and Resta[42] and implemented in VASP. The Bader charge, as an alternative approximation to $\mathbf{Z}^*$, is calculated by using the Bader charge analysis code[43-45].

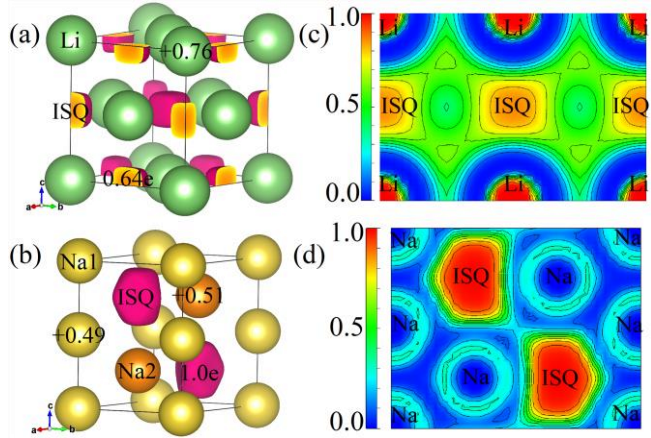

**FIG. 1. (Color online) Electron localization function (ELF) of (a) FCC-Li at 40 GPa and (b) hP4-Na at 300 GPa (isosurface=0.75), respectively. The charge state of atoms and ISQs are indicated. Note the prominent charge transfer and polarization. (c) and (d) show the ELF in (1 1 0) plane of (a) and (b), respectively.**

### III. Results

We first consider the limit case of $M_{ISQ} \to \infty$. In this case, the anionic ISQs are frozen and irresponsive to the vibrations of matrix atoms. Figures 2(a) and 2(b) shows the calculated $\omega(\mathbf{q})$ relation of isotropic dense FCC-Li at 40 GPa in this limit. The results represent a paradigm shift against the common wisdom about elemental metals. It is well-known that FCC structure has only one atom in the primitive cell, and all of such kinds of structure have only three acoustic branches, without any optical modes. Therefore it cannot have an LO-TO splitting even when there is a charge transfer from matrix atoms to the lattice interstices. However, we observe an energy splitting near the ZC (Γ point), totally out of expectation. Interestingly, it is a splitting occurs in the acoustic modes, rather than the expected optical modes. Namely, the long-range interaction in FCC-Li electride induces an anomalous LA-TA splitting, which is analog to the LO-TO splitting in ionic compounds. To understand this effect, we notice the introduction of localized electrons as a new species ISQ to the electride lattice, which renders the original acoustic modes the characteristics of an optical mode. In depth analysis reveals that it is the direct participation of electrons in lattice dynamics that leads to this acoustic branch splitting. We find that the two TA modes still have a null energy at Γ point, but the LA mode gains a finite energy and lifts up to a higher level at and near the Γ point. The frequency lift ($\Delta\omega=\omega_{LA}-\omega_{TA}$) is ranging from 140 to 430 cm$^{-1}$, according to whether the Born effective charge or the Bader charge are used. It is worth noting that the general feature of the splitting is robust, and is independent of the approximations that were used.

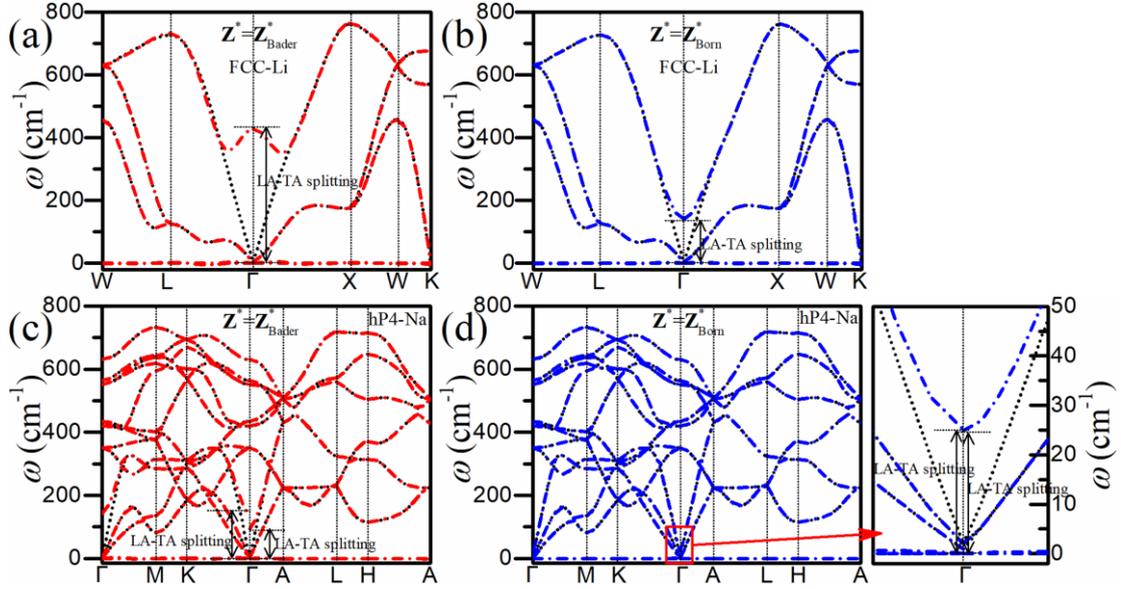

**FIG. 2. (Color online) Phonon dispersions of isotropic FCC-Li at 40 GPa (a,b) and anisotropic hP4-Na at 300 GPa (c,d) calculated using Bader charge (a,c) and Born effective charge (b,d) (short-dash-dotted lines), which are compared with the bare results that without taking the LRCI contribution into account (black short-dotted lines).**

The high-pressure phase of sodium (hP4-Na) is reported as a typical representative of anisotropic case (see Figs. 2(c) and 2(d)). Here the $\Delta\omega$ (25-132 cm$^{-1}$) is much smaller than that in FCC-Li. We note that the Bader charge of ISQ in hP4-Na is about 1.1$e$, comparing to 0.6$e$ in FCC-Li. The reduction of $\Delta\omega$ is mainly due to the larger dielectric constant in hP4-Na phase ($\varepsilon_{xx}^{\infty}=\varepsilon_{yy}^{\infty}=5.8, \varepsilon_{zz}^{\infty}=13.1$). For comparison, this value is $\varepsilon_{xx}^{\infty}=\varepsilon_{yy}^{\infty}=\varepsilon_{zz}^{\infty}=1.6$ in FCC-Li.

Structure of hP4-Na is unique because of the lattice anisotropy, which makes the LA-TA splitting also become anisotropic. As shown in Figs. 2(c) and 2(d), $\Delta\omega(\mathbf{q}\approx 0)|_{K\rightarrow\Gamma}$ is not equal to $\Delta\omega(\mathbf{q}\approx 0)|_{A\rightarrow\Gamma}$. It is necessary to point out that in Fig. 2(d) the non-analytical LRCI is almost suppressed when compared to Fig. 2(c). This is because the modern theory of polarization cannot capture the electron localization in the lattice interstice, and the resultant polarization properly. This leads to an underestimation of the charge state of the matrix atoms. In this regard the Bader charge (which is a scheme based on real space division) is more adequate to describe the charge state and the polarization in electrides. For example, the charge state of the matrix atom is about 0.76$e$ in Bader charge, compared to 0.20$e$ in Born effective charge in FCC-Li at 40 GPa, about 3.8 times larger. This charge difference linearly scales down the $\Delta\omega$ in Fig. 2 (see Fig. S1 in Supplementary material).

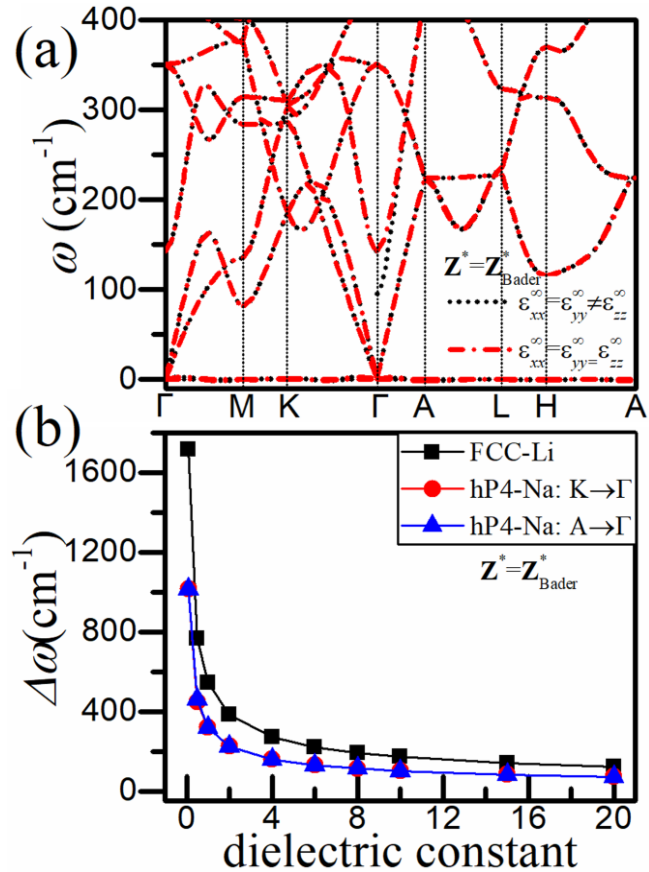

FIG. 3. (Color online) (a) Phonon dispersions of hP4-Na at 300 GPa calculated with artificially varied $\varepsilon^\infty$. (b) Variation of $\Delta\omega=\omega_{LA}-\omega_{TA}$ with $\varepsilon^\infty$ in FCC-Li at 40 GPa and hP4-Na at 300 GPa.

The influence of $\varepsilon^\infty$ is explored by artificially changes its value to investigate how the $\omega(\mathbf{q})$ curve is impacted. As shown in Fig. 3(a), when the $\varepsilon^\infty$ of hP4-Na are intentionally modified from $\varepsilon^\infty_{xx}=\varepsilon^\infty_{yy}=5.8, \varepsilon^\infty_{zz}=13.1$ to $\varepsilon^\infty_{xx}=\varepsilon^\infty_{yy}=\varepsilon^\infty_{zz}=5.8$, $\Delta\omega(\mathbf{q}\approx0)|_{K\to\Gamma}$ becomes equal to $\Delta\omega(\mathbf{q}\approx0)|_{A\to\Gamma}$, and the LA-TA splitting becomes isotropic. On the other hand, both $\Delta\omega$ for FCC-Li and hP4-Na show a sharp decrease with increasing dielectric constant, and $\Delta\omega$ approaches a small constant when $\varepsilon^\infty$ is bigger than 10 (see Fig. 3(b)). This not only reflects the magnitude of LA-TA splitting is inversely proportional to the square root of $\varepsilon^\infty$, as Eq.(2) indicated, but also reveals that $\varepsilon^\infty$ is an important source of the anisotropy. Please note that $\mathbf{Z}^*$ also contains anisotropy information, which however is missed if the Bader charge is used.

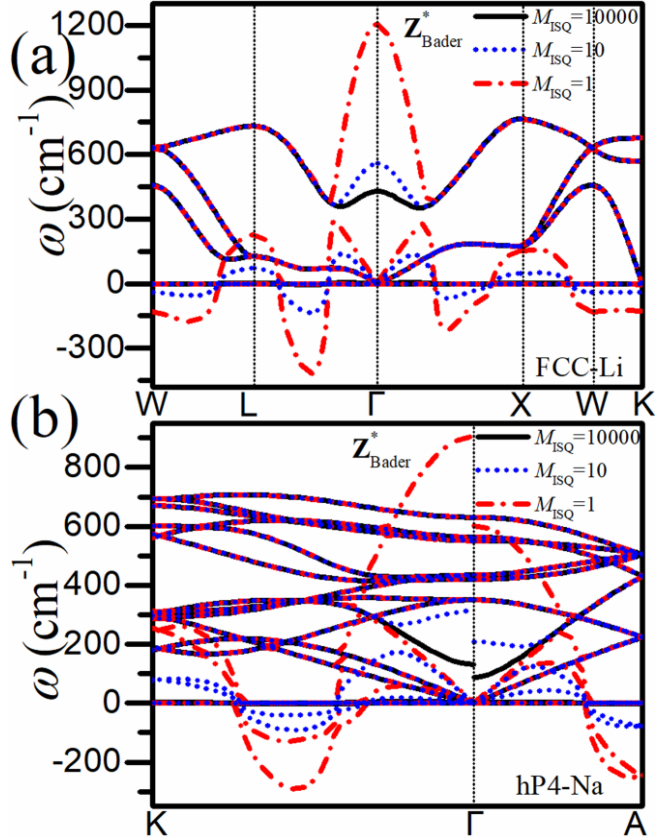

**FIG. 4. (Color online) Phonon dispersions of (a) FCC-Li at 40 GPa and (b) hP4-Na at 300 GPa calculated with different effective mass $M_{\text{ISQ}}$.**

In order to get an insight of the effect of the dynamic ISQ response, we lift the constraint of $M_{\text{ISQ}} \to \infty$. Figure 4(a) and 4(b) (see Figs. S2-S4 in the Supplementary material) display the variation of the $\omega(\mathbf{q})$ curves of FCC-Li and hP4-Na when $M_{\text{ISQ}}$ changed from 10000 to 1, respectively. It is evident that the movable ISQs provide an additional mode, which takes the place of the original LA branch, whereas the original LA mode moves up to become a special 'optical mode' that has relative displacements against the anionic ISQs. There are two band crossing-avoidances near the ZC, suggesting that both the up-moved 'optical mode' and the left behind LA mode have mixed contribution from both matrix atoms and ISQs. The fluctuating LA mode with small $M_{\text{ISQ}}$ implies strong energy dissipation by electrons when acoustic wave propagates through the electride. Nonetheless, since electrons have strong quantum characteristics, this classical mode of ISQ might be invalid when $M_{\text{ISQ}}$ is small. The important insight we obtained from this analysis is that the fictitious LA mode of electrons can be suppressed if ISQ is irresponsive to the lattice vibrations, and in this case the resultant LA-TA splitting has explicit physical implication. The up-lifted LA mode contains significant contributions from electrons. That is, it is the energy from the strong electron-phonon coupling that lifts the LA-TA degeneracy. It also reveals that if ISQs become responsive, the strong electron-phonon coupling will break the current standard theory of lattice dynamics, and the electrides must be treated with both electrons and matrix atoms on the same footing.

## IV. Discussion

By a thorough and comprehensive study of dense Li and Na, as well as many other promising electrides that are not reported here, we bring up with criteria (or rules of a thumb) for the LA-TA splitting in an electrode to occur: (i) The Bader charge of the ISQ must be greater than 0.2$e$; (ii) The $\varepsilon^{\infty}$ must not be bigger than 10, otherwise the LA-TA splitting could be too small to be observed.

We also noticed that in order to have a strong anisotropic LA-TA splitting, the electrode should have a very strong anisotropy in the dielectric tensor, or a strong anisotropy in $\mathbf{Z}^*$. It is worth noting that the dynamic response of ISQs to lattice vibration has a strong impact to the phonon dispersions. In the case when $M_{ISQ}\to\infty$ is valid, the LA-TA splitting is governed by Eq.(2), which can be applied to 0-D, 1-D, and 2-D electrides. For the case of a finite value of $M_{ISQ}$, the counterpart to Eqs.(3-4) for the 1-D and 2-D electrides can be straightforwardly derived following the similar reasoning. It is beyond the scope of this work, and we will not elaborate it here.

The non-analytic correction for the LRCI of Eqs.(2-4) is applicable as long as there are none zero polarization generated during lattice vibrations. In electride, this term becomes null only when ISQs response to the movement of matrix atoms in such a special manner that all involved atom-ISQ distances scaled linearly to cancel the generated dipoles. This subtle geometry balance is very rare, if not totally impossible. Such scaling correlation maintains the local relative positions of ISQ with respect to the matrix atoms, thus no LRCI occurs. We did not observe this kind of correlation in dense FCC-Li. Nonetheless, if it presents, this subtle balance can be removed by exerting an external magnetic field or electric field.

**V. Conclusion**

A theory was proposed to model the non-analytic contribution to the lattice phonon of LRCI that should present in electride. An anomalous LA-TA splitting was predicted in dense FCC-Li and

hP4-Na, and the direct effects of participation of localized electrons into lattice vibrational propagation as anionic ISQs were discovered. The strong electron(ISQ)-phonon coupling was revealed. After a thorough and extensive investigation of promising candidates of electrides, the criteria and the rules of a thumb for observable LA-TA splitting are summarized, which provides a guideline for experimentalists to design an appropriate experiment to detect this unconventional LA-TA splitting phenomenon. We also concludes that when electrons and ISQs are highly responsible to the lattice motion, the classical theory of lattice dynamics breaks, and the electrons and matrix atoms must be treated on the same footing with full quantum mechanics for this kind of mobile electrides. In practical applications, crystal defects, especially the potential colossal-charge-state impurities in electrides[46], may have striking influence on the local vibrational and optical properties of this class of emerging materials.

**Supplementary material**

See the Supplementary material for the information of the variation of $\Delta\omega$ with different Bader charge values and different $M_{ISQ}$.

**Acknowledgements**

This work was supported by the National Natural Science Foundation of China under Grant No. 11672274, the NSAF under Grant No. U1730248, the Science challenge Project under Grant No. TZ2016001, the CAEP Research Project CX2019002.

**Data availability**

The data that supports the findings of this study are available within the article and its Supporting information.

# Supplementary material for 'Prediction of anomalous LA-TA splitting in electrides'

Leilei Zhang,[1] Hua Y. Geng,[1,*] and Q. Wu[1]

[1]*National Key Laboratory of Shock Wave and Detonation Physics, Institute of Fluid Physics, CAEP, P.O. Box 919-102, Mianyang, Sichuan 621900, P.R. China*

## S1. Variation of Δω with different Bader charge values

Figure S1 shows the variation of the splitting magnitude $\Delta\omega=\omega_{LA}-\omega_{TA}$ with different Bader charge values in FCC-Li at 40 GPa and hP4-Na at 300 GPa. It gives a linear trend. It is worth noting that the $\Delta\omega$ for FCC-Li and hP4-Na are less than 200 cm$^{-1}$ when Bader charge value is smaller than 0.2$e$. Considering that the Bader charge is always greater than the Born effective charge (see Fig. 2 and 3 in the main text), in order to observe the noticeable LA-TA splitting experimentally, the Bader charge for ISQ should not less than 0.2$e$.

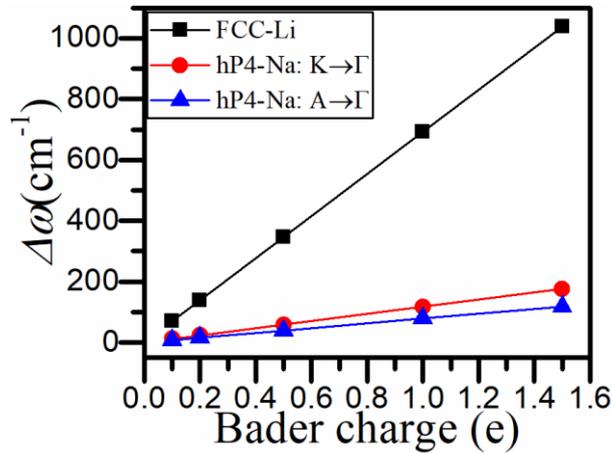

**Figure S1. (Color online)** Variation of $\Delta\omega=\omega_{LA}-\omega_{TA}$ with different Bader charge values in FCC-Li at 40 GPa and hP4-Na at 300 GPa.

## S2. Phonon dispersions of FCC-Li and hP4-Na with different $M_{ISQ}$

Figures S2 and S3 show the variation of the phonon dispersions of FCC-Li at 40 GPa and hP4-Na at 300 GPa when the effective mass of ISQs ($M_{ISQ}$) artificially changed from 1 to 10000, respectively. Please notice the band crossing avoidances around the zone center (Γ point) as the circle in Fig. S2(a) indicated, which are the fingerprint of hybridization of superimposition states in quantum mechanics. Here, one of the mixed states is from LA phonon mode, and another state is contributed from ISQs.

[*] To whom correspondence should be addressed. E-mail: s102genghy@caep.cn

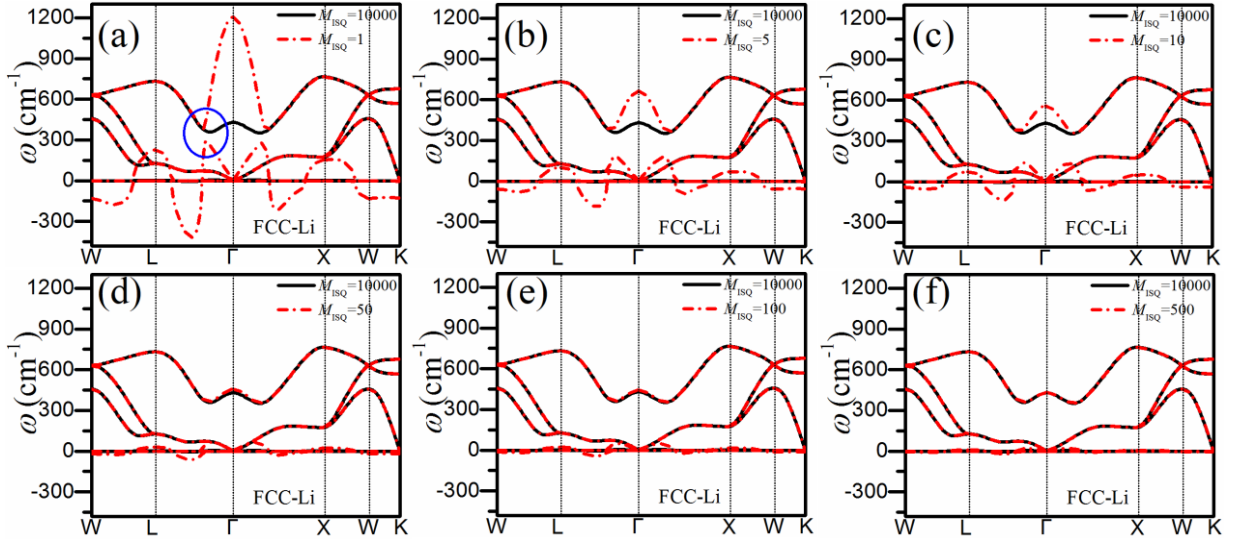

**Figure S2. (Color online) Variation of phonon dispersions of FCC-Li at 40 GPa with different value of $M_{ISQ}$.**

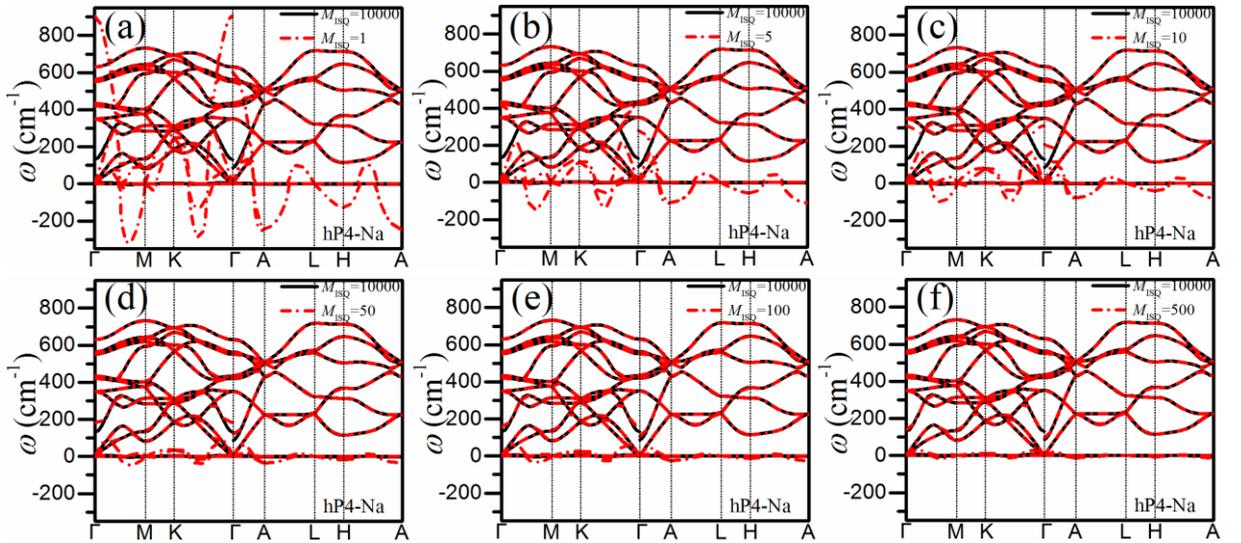

**Figure S3. (Color online) Variation of phonon dispersions of hP4-Na at 300 GPa with different value of $M_{ISQ}$.**

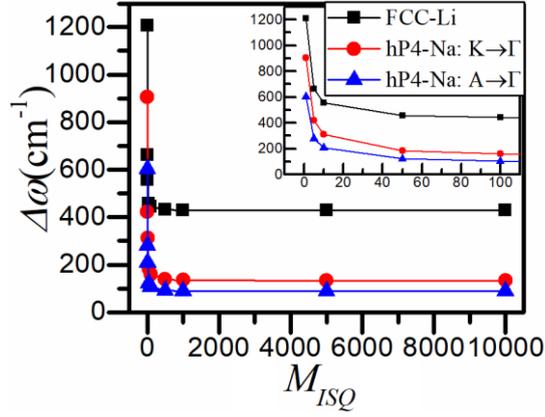

**Figure S4. (Color online) Variation of $\Delta\omega=\omega_{LA}-\omega_{TA}$ with different value of $M_{ISQ}$ in FCC-Li at 40 GPa and hP4-Na at 300 GPa. Inset shoes the detailed variation when $M_{ISQ}$ is less than 100.**

Figure S4 displays the variation of splitting magnitude $\Delta\omega=\omega_{LA}-\omega_{TA}$ with different value of $M_{ISQ}$ in FCC-Li at 40 GPa and hP4-Na at 300 GPa. It can be found that the $\Delta\omega$ of FCC-Li is always larger than that of hP4-Na. Please note that $\Delta\omega$ of FCC-Li and hP4-Na all approach a constant when $M_{ISQ}$ is bigger than 50.